\documentclass[%
 reprint,
 amsmath,amssymb,
 aps,
 prapplied,
 longbibliography,
]{revtex4-2}
\usepackage{pgf}

\usepackage{graphicx}
\usepackage{dcolumn}
\usepackage{bm}
\usepackage[mathlines]{lineno}
\usepackage{amsmath}

\begin{document}

\preprint{APS/123-QED}
    
\title{On-sky binary source hypothesis testing beyond the diffraction limit using spatial mode demultiplexing based detection}
\author{John S. Wallis}
\author{Ayden S. McCann} 
\author{Joshua J. Collier} 
\author{Lilani D. Toms-Hardman} 
\author{Alex M. Frost} 
\author{Benjamin P. Dix-Matthews} 
\author{David R. Gozzard}
\email{David.Gozzard@uwa.edu.au}
\altaffiliation[]{Department of Physics, The University of Western Australia, Crawley WA 6009, Australia}%
\affiliation{International Centre for Radio Astronomy Research, The University of Western Australia, Crawley WA 6009, Australia
}

\date{\today}

\begin{abstract}
   Improving the resolution of telescope systems will provide the opportunity to study new physical phenomena in previously unobserved environments. Spatial mode demultiplexing (SPADE) based imaging is a promising and rapidly evolving technique for improving the performance of optical telescopes beyond the diffraction limit. A key application of this technique is for near-optimal hypothesis testing for the presence of secondary and extended sources in the sub-diffraction regime. We present a proof-of-principle demonstration of a binary-SPADE based hypothesis testing instrument deployed on-sky. In our experiment, based on mode demultiplexing with a double-clad fiber coupler, we demonstrate detection of a binary star system separated below the diffraction limit, using a system with an 11.8~mm aperture, $D/r_0 = $ $0.27$ and a 10~nm bandpass filter. We find the scaling of the system's type II error rate (the probability of missing a binary) is heavily limited by unbalanced loss in our double-clad fiber coupler when compared to the idealized quantum limits. The measured error rates are well described by a model that includes atmospheric turbulence and loss. They outperform simulated direct imaging under similar experimental conditions, but remain well above both the quantum limit and ideal direct imaging. We expect that if this instrument is scaled to larger aperture telescope systems the effects of atmospheric turbulence will degrade this system's performance and that outperforming direct imaging in binary hypothesis testing will necessitate a switch to a lower-loss mode sorting technology.
\end{abstract}

    \maketitle


\section{Introduction}\label{sec:intro}
The astronomy community continues to demand greater angular resolution from telescope systems. To this end, optical astronomy research has investigated techniques to extract more information from single aperture telescopes. Using methods borrowed from quantum information theory, significant improvements to the performance of optical telescopes were theorized to be obtainable by replacing direct imaging (DI) of photons with a new family of quantum optimal measurements \cite{TsangSuperresolution, LuHypothesisTesting}. A practical technique to implement these measurements has been coined SPAtial mode DEmultiplexing (SPADE), which decomposes the incoming field into a carefully chosen set of spatial modes. When compared to measuring in the ``pixel'' modes of direct imaging these alternative spatial modes allow for more spatial information to be extracted from each detected photon. Experimental studies have demonstrated the performance of SPADE using different mode projection technologies including multi-plane light converters (MPLC) \cite{BoucherMPLC,SantamariaMPLC,RouviereMPLC,TanMPLC, WallisMPLC, MokuNN, LSantamaria:2023, AntoninTwoMPLCS, BaesianMPLC}, photonic lanterns \cite{OnSkyPL, NorrisPL}, spatial light modulators \cite{PuskhinaSLM} and shaped local oscillator based interferometric techniques \cite{FYangLO}. The performance of this method has been studied in the context of quantum-limited source parameter estimation \cite{WallisMPLC} and hypothesis testing \cite{LuHypothesisTesting, NoiselessHT, CrossTalkHT, PracticalHT, WallisDCF}. In Tsang's original proposal \cite{TsangSuperresolution}, a simple method to implement SPADE with the modes of an optical waveguide was proposed. In this article, we use a double-clad fiber (DCF) as a simple binary mode sorter in order to implement a binary-SPADE hypothesis testing protocol. MPLC and photonic-lantern technologies are the most immediately applicable for astronomical instruments, as SPADE-type local-oscillator-based measurements do not allow for optimal detection of broadband incoherent sources. While DCF can also mode-sort incoherent light, it only allows for a binary mode sorting between a fundamental Gaussian mode and all other higher order modes. Due to the large number of reflections in MPLCs, extremely strict alignment is required for accurate imaging \cite{BoucherMPLC, WallisMPLC}. DCF is no harder to align than a single-mode fiber, for which the astronomy community has developed methods for efficient coupling of stellar sources \cite{SRfibrecoupling} and, for measurements only concerned with binary mode sorting, the lower cost and lower complexity make it a sensible choice as a mode sorter. Binary-SPADE systems would supplement existing observatories and could aid in identification of binary systems and exoplanets \cite{NoiselessHT}. For more complex sources binary-SPADE would be useful in testing higher order models. The discovery that a previously identified X-ray black hole binary is a triple system \cite{BlackholeDiscovery} is an example of one such application where binary-SPADE could be useful.

While SPADE-based measurements have been well studied in laboratory environments \cite{BoucherMPLC, SantamariaMPLC, RouviereMPLC, WallisDCF, WallisMPLC, PuskhinaSLM, FYangLO}, to our knowledge only the Fibered Imager foR a Single Telescope--Photonic Lantern (FIRST-PL) instrument on the Subaru telescope has demonstrated on-sky performance \cite{OnSkyPL} of a SPADE-based measurement. The FIRST-PL instrument used a photonic lantern to measure sub-diffraction limited spectroscopic photocenter locations. A key problem for any sub-diffraction imaging is first determining when the subject of an image consists of more than just a single point source. We present on-sky measurements using SPADE for sub-diffraction limited source classification. Previous demonstrations of SPADE-based hypothesis testing were made in the laboratory \cite{WallisDCF, BaesianMPLC}; here we apply it to astronomical sources, using a DCF-based instrument. We focus our demonstration on measurements containing fewer than 5000 photons, a regime where the per-photon statistical power should be maximized, for which SPADE is well suited. 

In astronomical contexts where binaries and extended sources are relatively rare compared to single point sources (i.e., stars), this problem is often treated as a form of an asymmetric hypothesis test. In the formalism of quantum information theory, the telescope instrument samples photons in the image plane with an assumed photon distribution calculated by the instrument's point spread function (PSF). In our experiment we assume the null hypothesis $H_0$ is the distribution which arises from sampling a single point source \cite{NoiselessHT}. As such we can model the distributions of photon positions that DI would measure with the random variable given by position probability density:
\begin{equation}
    p_0(x) = |\psi_{psf}(x)|^2,
\end{equation}

where $\psi_{psf}(x)$ is the instrument's PSF. The alternative hypothesis $H_1$ is that the object of interest is produced by two point sources with a relative brightness $\epsilon$ and an angular separation of $\theta$, i.e. a separation $r$ in the image plane. If we work in a frame of reference where $x$ = 0 corresponds to the image's optical center of mass then the photon distribution under such a hypothesis can be written as: 
\begin{equation}
p_1(x) = \epsilon|\psi_{psf}(x - (1-\epsilon)r)|^2 + (1-\epsilon)|\psi_{psf}(x + \epsilon r)|^2.
\end{equation}
\section{Instrument and model}\label{sec:instrument}
\begin{figure}
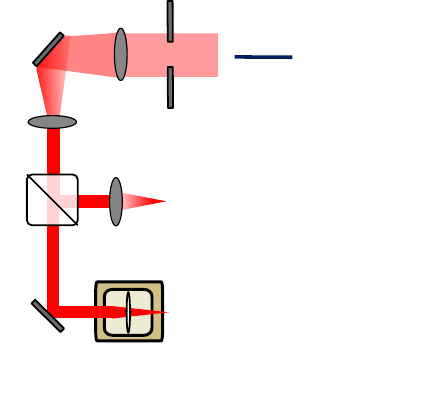
\caption{\label{fig:insturmentSchema} Optical instrument for on-sky hypothesis testing. L1: f=150 mm mode matching lens. L2: f=-20 mm mode matching lens. L3: f=30 mm imaging lens. L4: f=12.19 mm fiber coupling lens. BS: 90:10 beam splitter. F1: 780 $\pm$ 5 nm optical filter. CAM: camera for pointing model and tracking. IRIS: 11.8 mm aperture stop to set the point spread of the imaging system. DCF-C: double-clad fiber coupler used to separate the fundamental and higher order modes. SP-APD: silicon single-photon avalanche photodiodes used to monitor photons in the fundamental and higher order modes.}
\end{figure}

To collect on-sky observations we designed the instrument shown in Fig. \ref{fig:insturmentSchema}. The system has a primary aperture diameter $D=11.8$~mm and an effective focal length of EFL =~$90$~mm on the DCF path. We used a pair of silicon-based avalanche photodiodes, which have a specified dark count rate of 100 cps and a detection efficiency of $>70\%$ at $\lambda=780$~nm. They were connected to a time-tagger to measure photon arrivals. We limited the observation wavelength to $\lambda=780\pm 5$~nm which corresponds to the specified cut-off wavelength of the DCF. Using simulations, we chose optics to optimize the system's point spread function to match the fiber mode field diameter. Simulations predicted a coupling loss of 2.4~dB. To align the instrument, we launched light from the single-mode DCF port towards a corner-cube reflector placed horizontally $1.2$~km away on a neighboring building. The measured return power indicated a coupling loss of 7.0~dB. This loss includes fiber coupling, turbulence and contaminants on the corner cube. 

A DCF consists of a single-mode core surrounded by a multi-mode inner cladding and a lower-index outer cladding. We use a tapered DCF coupler (Thorlabs DC780SE2FA), in which the 4~$\mu$m core (NA 0.13) guides only the fundamental LP$_{01}$ mode at 780~nm, while the 23~$\mu$m inner cladding (NA 0.2) guides the higher-order modes, giving 180 guided modes in total. In the coupler the DCF is fused to a 200/220~$\mu$m multi-mode fiber and tapered. The fundamental mode passes through the taper to the single-mode port, which is connected to single-mode fiber. Cladding modes that are cut off in the narrow waist of the taper spread into the fused cladding and are captured by the multi-mode fiber, which forms the multi-mode port, while cladding modes that remain in the DCF inner cladding are lost in the single-mode fiber spliced to the coupler. The coupler therefore acts as a binary mode sorter, separating the fundamental LP mode from the higher-order modes.

To perform a binary-SPADE hypothesis test we use the DCF to transform photon position statistics into photon mode statistics. If we design the optical system to maximize the overlap of the system's PSF with the fundamental mode of the DCF, then we can monitor the single (fundamental) mode port and the multi-mode port with single-photon detectors. Throughout this analysis and experiment, we work in a regime where detector dead time is insignificant compared to the photon arrival rate. Under these conditions each source couples incoherently into the DCF's single-mode port with probability based on the mode overlap with the diffracted field on the image plane.
The guided DCF modes $\phi_\mu(\mathbf{r})$, $\mu=1,\dots,M$, are obtained by simulating the guided modes of the fiber ($M=180$ for our DCF). A source at position $\mathbf{r}_i$ in the image plane couples into mode $\mu$ with probability
\begin{equation}
\eta_\mu(\mathbf{r}_i)=\frac{\left|\int\mathrm{d}^2r\,\phi^*_\mu(\mathbf{r})\psi_{psf}(\mathbf{r}-\mathbf{r}_i)\right|^2}{\int\mathrm{d}^2r|\phi_\mu(\mathbf{r})|^2\int\mathrm{d}^2r|\psi_{psf}(\mathbf{r})|^2},
\label{eq:Psm}
\end{equation}
so that under hypothesis $H_j$ the fractions of the light coupled into the single-mode and multi-mode ports are
\begin{equation}
P_{\mathrm{opt}}^{(j)}=\sum_i\epsilon_i\,\eta_1(\mathbf{r}_i),\qquad Q_{\mathrm{opt}}^{(j)}=\sum_i\epsilon_i\sum_{\mu=1}^{M}c_\mu\eta_\mu(\mathbf{r}_i),
\label{eq:Popt}
\end{equation}
where $c_\mu$ is the DCF's multi-mode extraction efficiency ($c_1$ being the coupler's crosstalk from the fundamental mode into the multi-mode port), $\epsilon_i$ is the $i$th source's relative brightness and $\mathbf{r}_i$ its position under hypothesis $H_j$. As the guided modes are orthogonal, $P_{\mathrm{opt}}^{(j)}+Q_{\mathrm{opt}}^{(j)}\le 1$, with the remainder of the light not guided by the fiber. We assume that multi-mode light exiting the DCF on the single-mode port is lost in the 1~m of single-mode fiber on its way to the detector.

To model the effects of turbulence, we adopt the modeling framework from \cite{SRfibrecoupling} assuming no rapid image stabilization. The coupling efficiency into the DCF fundamental mode is averaged over the random phase of the incoming wavefront, i.e. the average is taken over the overlap $\eta_\mu(\mathbf{r}_i)$ in Eq.~(\ref{eq:Psm}). Writing this overlap in the pupil plane, a source at $\mathbf{r}_i$ couples into mode $\mu$ with an average probability
\begin{multline}
\langle\eta_\mu(\mathbf{r}_i)\rangle=\frac{1}{\mathcal{N}_\mu}\iint\mathrm{d}^2u\,\mathrm{d}^2u'\,G_\mu(\mathbf{u})G_\mu^*(\mathbf{u}')\\
\times B(\Delta\mathbf{u})\,e^{-ik\Delta\mathbf{u}\cdot\mathbf{r}_i/\mathrm{EFL}},
\label{eq:SR}
\end{multline}
where $G_\mu(\mathbf{u})=\Psi_{psf}(\mathbf{u})\Phi^*_\mu(\mathbf{u})$, with $\Psi_{psf}$ and $\Phi_\mu$ the pupil-plane (Fourier) representations of $\psi_{psf}$ and $\phi_\mu$, $\mathbf{u}$ is the pupil coordinate, $\Delta\mathbf{u}=\mathbf{u}-\mathbf{u}'$, $k=2\pi/\lambda$, $\mathcal{N}_\mu=\int\mathrm{d}^2u|\Phi_\mu(\mathbf{u})|^2\int\mathrm{d}^2u|\Psi_{psf}(\mathbf{u})|^2$ and
\begin{equation}
B(\Delta\mathbf{u})=\exp\left(-3.44\left(\frac{|\Delta\mathbf{u}|}{r_0}\right)^{\frac{5}{3}}\right)
\end{equation}
is the long-exposure coherence of the wavefront between two points in the pupil for Kolmogorov turbulence, where $r_0$ is the Fried parameter. Turbulence reduces the coherence between separated parts of the pupil, which lowers the coupling into the fundamental mode and increases the coupling to the multi-mode port. We replace each overlap in Eq.~(\ref{eq:Popt}) with $\langle\eta_\mu(\mathbf{r}_i)\rangle$, so that a binary contributes the sum of the averaged couplings of its two stars weighted by $\epsilon_i$. This leaves the Fried parameter $r_0$ as the only turbulence parameter in the model. We determine $r_0$ from the fluctuations of the coupling on millisecond timescales (Appendix~\ref{app:sr}), which gives $r_0=4.4\pm0.8$~cm at zenith. We note that this describes the average coupling over many realizations of the turbulence, and so does not capture the correlation between photons that arrive within the same atmospheric coherence time.

Table~\ref{table:modes} gives the simulated coupling $\eta_\mu$ from Eq.~(\ref{eq:Psm}) into the modes of the DCF for a point source ($H_0$) and for Alpha Centauri ($H_1$). For a point source 58\% of the light couples into the fundamental mode and 30\% into LP$_{03}$, whose radial nodes overlap the first Airy ring. The binary nature of Alpha Centauri lowers $\eta_1$ from 0.575 to 0.477 and excites the antisymmetric LP$_{11}$ mode.

The extraction efficiency $c_\mu$ depends on whether a mode is guided in the tapered waist of the coupler. We simulate a fused, tapered coupler between the DCF and a 200/220~$\mu$m multi-mode fiber with a beam-propagation method, launching each guided mode of the DCF. For a tapered coupler with the best combined single-mode and multi-mode efficiency, the modes up to LP$_{03}$ remain guided, and are then lost in the single-mode fiber, while the higher order modes near cut-off are partially extracted into the multi-mode fiber. These modes carry most of the information in the multi-mode port. The simulated coupler transmits 98\% of the fundamental mode to the single-mode port and extracts almost 50\% of the light under overfilled launch conditions \cite{LemireRenaudDCFC}.

\begin{table}
\centering
\begin{ruledtabular}
\begin{tabular}{cccc}
Mode  & $\eta_\mu$ ($H_0$) & $\eta_\mu$ ($H_1$) & $c_\mu$\\
\hline
LP$_{01}$  & 0.575 & 0.477 & 0.006\\
LP$_{11}$  & 0 & 0.110 & 0.002\\
LP$_{02}$  & 0.014 & 0.025 & 0.004\\
LP$_{21}$  & 0 & 0.010 & 0.003\\
LP$_{03}$  & 0.295 & 0.236 & 0.012\\
LP$_{13}$  & 0 & 0.019 & 0.008\\
LP$_{04}$  & 0.013 & 0.010 & 0.068\\
LP$_{05}$  & 0.010 & 0.008 & 0.491\\
other  & 0.000 & 0.007 & $<0.5^\dagger$\\
not guided & 0.093 & 0.097 & --\\
\end{tabular}
\end{ruledtabular}
\caption{Simulated coupling $\eta_\mu$ for the guided LP mode families of the DCF for a point source ($H_0$) and Alpha Centauri ($H_1$, $\epsilon=0.26$, 9.40$''$), without turbulence, and the simulated multi-mode extraction efficiency $c_\mu$ of a coupler with a taper ratio of 0.55 and a 10~mm hot zone ($c_1$ is the crosstalk from the fundamental mode). Values are summed over the degenerate modes of each family. $^\dagger$ averaged over the rest of the 180 supported LP modes.}
\label{table:modes}
\end{table}
For on-sky observations we generate a series of measurements, each with a fixed number of total detected photons $N$. Then by treating the total number of detected single-mode photons as a binomial distribution of $N$ events,  the likelihood ratio test statistic $\Lambda_t$ for $n$ single-mode photons is given by:

\begin{equation}
  \Lambda_t = -2\ln\left[\frac{\mathcal{L}(H_0|n,N)}{\mathcal{L}(H_1|n,N)}\right],
\end{equation}
with likelihoods calculated from the fact $n\sim \mathrm{Bin}(N,P_{\mathrm{sm}})$ with $P_{\mathrm{sm}}$ calculated from Eq.~(\ref{eq:Psmdet}) for hypotheses $H_0$ and $H_1$.

The photon statistics therefore follow

\begin{multline}
P(n \mid H_j, N)
=
\binom{N}{n}
\left(P_{\mathrm{sm}}^{(j)}\right)^n
\left(1-P_{\mathrm{sm}}^{(j)}\right)^m,
\\
\qquad
N=n+m,\quad j\in\{0,1\},
\end{multline}

with $P_{\mathrm{sm}}$ given by

\begin{equation}
P_{\mathrm{sm}}^{(j)}
=
\frac{\left(P_{\mathrm{opt}}^{(j)}+\delta^{(j)}\right)}
{\left(P_{\mathrm{opt}}^{(j)}+\delta^{(j)}\right)+\left(Q_{\mathrm{opt}}^{(j)}+\delta^{(j)}\right)},
\label{eq:Psmdet}
\end{equation}
where $n$ and $m$ are the number of photons detected in the single-mode and multi-mode ports respectively. Here $\delta^{(j)}=d/\Phi^{(j)}$ is the ratio of the specified dark count rate of each detector, $d=100$~cps, to the source photon rate at the DCF, $\Phi^{(j)}=R/(P_{\mathrm{opt}}^{(j)}+Q_{\mathrm{opt}}^{(j)})$, where $R$ is the total detected count rate (Table~\ref{table:obs}). Dark counts are independent Poisson events in each port, so $n$ remains binomially distributed given $N$; for Alpha Centauri $\delta^{(j)}\approx3\times10^{-3}$.

We note that this analysis ignores the possibility of an extended source. We leave a detailed analysis of quantitative approaches to testing for such sources to future work. However, because an asymmetric hypothesis test can only reject the null hypothesis, and the photon distribution of an extended source should differ significantly from that of a point source, a suitable test will reject $H_0$, so our technique will be immediately applicable for testing for the presence of extended sources.

\section{On-sky observations}\label{sec:observations}After calibration and alignment the system was placed on a telescope mount and used to collect SPADE photon counts in on-sky observations. Observations were taken of Alpha Centauri A and B, Sirius, Betelgeuse, Canopus and Jupiter on the nights of the 14th and 18th of March 2026. During this experiment the instrument was manually aligned to optimize the single-mode detection rate and then the targets were observed for 100 seconds using open-loop tracking on a motorized mount (PlaneWave L-350). The pointing model was acquired through the camera with a residual rms error of $<1''$. During this time period the single-mode and multi-mode photon events were digitally time-tagged. Average optical throughput, mode isolation and observation altitude are shown in Table~\ref{table:obs}. We observed Jupiter to better understand the response of our system to larger extended sources. The DCF decisively rejects Jupiter as a point source. As it is not a sub-diffraction binary, we exclude it from further analysis.

\begin{table}
\centering
\begin{ruledtabular}
\begin{tabular}{cccc}
Name & Altitude (deg) &Photons/s & Isolation (dB) \\
\hline
Canopus & 54.42 & 22668 & 16.12 \\
Sirius & 54.30 & 32477 & 16.10 \\
Betelgeuse & 32.36 & 36298 & 16.00 \\
Alpha Centauri & 32.73 & 19347 & 15.45 \\
Jupiter & 27.81 & 32246 & 10.27 \\
\end{tabular}
\end{ruledtabular}
\caption{Detected optical throughput (photons/s), isolation (dB) and altitude (deg) at observation time. Jupiter is a large extended source and as such the instrument observes a lower SM isolation. The effective point sources Canopus, Sirius and Betelgeuse all have higher SM isolations than the binary system Alpha Centauri.}
\label{table:obs}
\end{table}

Several technical details of the operation of the DCF heavily limit the performance of the system. Because a diffraction-limited Airy PSF can overlap the fundamental mode by at most 81$\%$, there is already an intrinsic crosstalk associated with demultiplexing light diffracted by a circular aperture into fiber LP modes. The excited higher-order modes have the majority of their power concentrated in the next four modes. We characterized the performance of the coupler in extracting multi-mode light under these conditions on an optical table and measured an additional loss of 11.54~dB, which is due to the low extraction efficiency $c_\mu$ of the lower-order modes in the tapered coupler model. The excess loss and mode mismatch heavily limited the distinguishability between point and extended sources. Atmospheric turbulence also limits the performance of the system.

\section{Hypothesis-testing results}\label{sec:results}From the time-tagged photons we can evaluate the performance of our system during binary-source hypothesis testing. Because binaries are somewhat rare in astronomy it is useful to use an asymmetric hypothesis test to reject the null hypothesis that the image is a single point source. We fix the chance to falsely reject the null hypothesis of all tests in this paper to $\alpha\le0.05$, and then evaluate the chance to incorrectly assume the null hypothesis in the case of a binary, i.e. the type II error rate $\beta$. We study $\beta$ as a function of the number of observed photons $N$; to do so, we group the photon events into bins of size $N$. Then we assume that Sirius, Betelgeuse and Canopus are effectively point sources, calculate $\Lambda_t$ for each measurement and then choose the smallest threshold $\tau_N$ for which at most 5\% of the measurements of each of these stars have $\Lambda_t> \tau_N$. As $\Lambda_t$ only takes discrete values, we use a randomized threshold, rejecting $H_0$ with a fixed probability when $\Lambda_t=\tau_N$, so that $\alpha=0.05$ exactly for the least favorable point source. Appendix \ref{app:errors} gives a detailed definition of the exact distributions of $\alpha$ and $\beta$ under this method. To avoid calibrating and evaluating the test on the same data, we choose $\tau_N$ using one half of each observation and evaluate $\alpha$ and $\beta$ on the other half, and then swap the two halves. Note that while Sirius is a binary, the larger difference in brightness and small angular separation mean that, from the point of view of our instrument, it is effectively a point source. We then calculate $\Lambda_t$ when observing Alpha Centauri. The type II error ($\beta$) is the proportion of these measurements for which $H_0$ is not rejected, i.e. with $\Lambda_t \le \tau_N$ (with the randomized rejection applied at $\Lambda_t=\tau_N$). Over all point source measurements the evaluated $\alpha$ lies between 0.034 and 0.050.  We repeat this procedure for $N$ ranging from 10 to 5000 with our data set; this calibration is done each night and the evaluated measurements from both nights are then pooled. For $N=10$, $\beta$ is evaluated with a total of 1900000 and 400000 single and double source measurements, respectively. When binning for $N=5000$ this decreases to 3800 and 800 respective measurements. Measurements that are close in time are correlated by atmospheric turbulence and by slow drifts in alignment, so we estimate 95\% confidence intervals for $\beta$ by shuffling the sample order of 5~s blocks of data within each observation.

\begin{figure}
\input{beta_value_comparison_semilogx_r0glrt.pgf}
\caption[Type II error rate against photons per measurement]{\label{fig:results} Type II error rate ($\beta$) against the photons per measurement $N$ for Alpha Centauri, with $\alpha\le0.05$. Experimental data (blue points) are the on-sky $\beta$ from both nights, using a randomized likelihood ratio threshold $\tau_N$ that is calibrated on the point sources to keep $\alpha\le0.05$. The light blue region is the 95\% confidence interval. The red curve is the modeled DCF performance using the simulated coupler of Table~\ref{table:modes} and the turbulence model of Eq.~(\ref{eq:SR}) with $r_0$ determined from the coupling fluctuations (Appendix~\ref{app:sr}). Green shows the quantum limit for binary SPADE, a mode sorter matched to the system's PSF with no turbulence, noise or loss. Orange shows ideal DI with a noiseless detector and no turbulence. Purple shows simulated DI of Alpha Centauri (separation = 9.4$''$, $\epsilon = 0.26$) with a $64\times64$ array of the same single-photon detectors used in our instrument under comparable turbulence (zenith $D/r_0=0.27$), using a generalized likelihood ratio test in which the binary parameters and $r_0$ are unknown. For the simulated DI, $N$ is the number of detected events, including dark counts only from pixels used in the hypothesis test.}
\end{figure}

Figure \ref{fig:results} shows the evaluated error rates $\beta$ as we increase the number of photons per measurement $N$ when using the on-sky calibration. We plot the evaluation of the on-sky data in blue.  We plot in red the simulated performance of a system with the same PSF as ours accounting for the modeled LP mode couplings and MM extraction loss.  The modeled red curve matches closely to our measured performance. A zenith Fried parameter of 4.4~cm is indicative of extremely strong turbulence, but is reasonable given that the observations were taken on the roof of the UWA physics building, at an urban university in a low-elevation coastal city, on warm nights in early autumn, and that this site experiences strong turbulence induced by nearby building vents. In future work we plan to evaluate the performance of this system for $r_0>10$~cm at $780$~nm, which is typical of good astronomical sites. As a point of comparison, we simulated ideal DI's performance for the same photon number $N$ with a noiseless $100\times100$ array of pixels that only experience shot noise. We ignore the effects of dark noise and read noise. The simulated DI measurement had a perfect matched filter with prior knowledge of Alpha Centauri's separation. We also assumed DI had a diffraction limited PSF for an aperture size equal to our instrument and that the simulated  DI experienced no turbulence. Under these ideal conditions DI reaches $\beta=0.5$ at $N\approx60$, while the quantum limit for binary SPADE, a mode sorter matched to our PSF with no turbulence, noise or loss, reaches $\beta=0.5$ at $N\approx8$. Our measured $\beta$ remains well above both, as our instrument is limited by the excess loss of the coupler and by turbulence. We also simulated a more realistic DI measurement, with a $64\times64$ array of the same single photon detectors used in our instrument (100~cps dark counts per pixel) under comparable turbulence ($D/r_0=$ $0.27$) and with a 1$''$ rms pointing jitter, with the binary parameters and $r_0$ unknown. This simulated DI measurement has a higher $\beta$ than our measured $\beta$ at every $N$ we evaluated.

While the various sources of loss heavily limited the performance of this system, the technique is expected to work with lower-loss mode demultiplexers. In Sec.~\ref{sec:multiport} we discuss the simulated performance of currently achievable demultiplexers on state-of-the-art telescopes with realistic residuals from extreme adaptive optics systems.
\section{Expected performance on state-of-the-art telescopes}\label{sec:multiport}The DCF used on sky splits the light into only two channels, and its crosstalk limits its performance. Here we model two receivers that resolve the light into six spatial modes, each routed to its own single-photon detector, on a $D=10$~m telescope at $\lambda=1550$~nm with adaptive optics (AO). The first is a six-port mode-selective photonic lantern (PL) \cite{XinPLN}, which maps the six spatial modes of the LP$_{01}$, LP$_{02}$, LP$_{11}$ and LP$_{21}$ families onto separate single-mode outputs. The second is a six-mode multi-plane light converter (MPLC) \cite{BoucherMPLC}, which projects onto the first six Hermite--Gauss modes. Each port is coupled with probability $\eta_\mu$ as in Eq.~(\ref{eq:Psm}), and both receivers have $-30$~dB crosstalk between their ports. We take $r_0=75$~cm ($D/r_0=13.3$) and use the results of \cite{RigautAO, JolissaintAO} to model such an AO system with a Strehl ratio of 0.90. For each measurement of $N_{\mathrm{det}}$ detected photons we apply the likelihood-ratio test to the port counts $n_\mu$,
\begin{equation}
\Lambda_t=2\sum_{\mu=1}^{6}n_\mu\ln\frac{p_\mu^{(1)}}{p_\mu^{(0)}},\qquad
p_\mu^{(j)}=\frac{\eta_\mu^{(j)}}{\sum_\nu\eta_\nu^{(j)}},
\label{eq:app_lrt}
\end{equation}
where $\eta_\mu^{(j)}$ is the modeled overlap for each mode. Undetected light carries no information, so we convert $N_{\mathrm{det}}$ to the number of photons $N$ entering the receiver by dividing by the fraction of a single star's light that is detected. We compare with the quantum limit \cite{LuHypothesisTesting, NoiselessHT}, reached by binary SPADE with a PSF-matched sorter and no turbulence,
\begin{equation}
\beta^{\mathrm{QL}}=(1-\alpha)(1-q)^N,
\label{eq:app_ql}
\end{equation}
where $q=1-\sum_i\epsilon_i\left|\langle\psi_{psf}(\mathbf{r})|\psi_{psf}(\mathbf{r}-\mathbf{r}_i)\rangle\right|^2$ is the probability that a photon from the binary is orthogonal to the PSF mode, and with ideal DI: a noiseless photon-counting camera with $\lambda/8D$ pixels and no turbulence. While these receivers have more modes than true binary SPADE, the principle of operation is similar: some modes are nulled when the receiver is pointed at a single point source, which gives high sensitivity to sub-diffraction binaries.

\begin{figure}[t]
\includegraphics[width=\columnwidth]{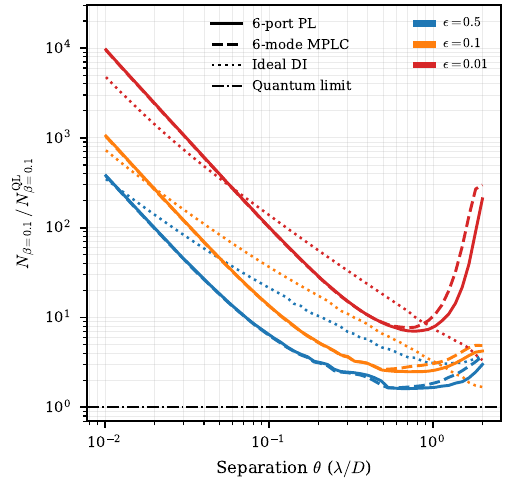}
\caption{Photons needed to reach $\beta=0.1$ ($\alpha=0.05$), relative to the quantum limit, against binary separation for the six-port photonic lantern (solid) and the six-mode MPLC (dashed) with $-30$~dB crosstalk, both under AO with a Strehl ratio of 0.90 ($D=10$~m, $\lambda=1550$~nm, $r_0=75$~cm). Dotted: ideal DI without turbulence. Colors give the companion's relative brightness $\epsilon$.}
\label{fig:app_lantern_mplc}
\end{figure}

Figure~\ref{fig:app_lantern_mplc} shows that the two receivers perform almost identically below $\theta\approx0.5\,\lambda/D$. Without turbulence (not shown) both reach $1.41$ times the quantum-limited photon number at small separations. This floor is set by the coupling $\eta=0.71$ of the PSF's first-order (tilt-like) mode, $\partial_x\psi_{psf}$, captured by the LP$_{11}$ or HG$_{10}$ port, which carries the companion's signal when $\theta\ll\lambda/D$; LP$_{11}$ and HG$_{10}$ are nearly the same mode. Background light left uncorrected by the AO dominates for faint and close companions, so the penalty grows as $\theta$ decreases. At $\theta=0.05\,\lambda/D$ the receivers need 18, 46 and 390 times the quantum limit for $\epsilon=0.5$, 0.1 and 0.01. Both still beat turbulence-free ideal DI from $\theta\approx0.012$, 0.02 and $0.06\,\lambda/D$ for $\epsilon=0.5$, 0.1 and 0.01, needing up to 3.6, 3.3 and 2.2 times fewer photons near $\theta\approx0.1$--$0.4\,\lambda/D$. At smaller separations ideal DI is better; at $\theta=0.01\,\lambda/D$ it needs 0.92, 0.69 and 0.50 times the photons of the receivers. The performance degradation at lower values of $\epsilon$ is set by the AO residuals and the demultiplexer crosstalk. Mode-selective photonic lanterns have been demonstrated with fundamental mode isolations approaching $80$~dB \cite{BecerraDeanaDCFLantern}, whereas MPLCs have yet to reach such low crosstalk. MPLC demultiplexing performance is also sensitive to alignment, so photonic lanterns are likely to be the better choice for high-performance hypothesis testing under realistic conditions. Beyond $\theta\approx\lambda/D$ the faint companion leaves the effective field of view of the demultiplexers, and for $\epsilon\le0.1$ ideal DI overtakes both receivers at $\theta\approx0.9$--$1.2\,\lambda/D$. Adding more modes increases the field of view, but direct imaging is close to optimal once the sources are separated by a few times more than the diffraction limit. SPADE-based hypothesis testing is therefore a poor choice for faint companions separated by more than the diffraction limit.

\section{Conclusion}\label{sec:conclusion}In conclusion, we have demonstrated a proof-of-principle instrument for binary source classification and applied the binary-SPADE technique to on-sky observations. This system was limited by strong turbulence and poor demultiplexing loss performance but follows the expected SPADE error scaling under these conditions. The measured error rates remain well above both the quantum limit and ideal direct imaging, and indicate that the technical limitations of the demultiplexing loss and crosstalk of the DCF make it unsuitable for high-performance SPADE measurements. This work paves a path for SPADE systems with greater mode counts and better loss performance based on MPLC or photonic-lantern technology, which can not only perform source hypothesis testing but also parameter estimation  on sub-diffraction sources. Future work will also involve testing of larger apertures at sites with better astronomical seeing.

\begin{acknowledgments}
J.S.W., J.J.C., A.S.M. and A.M.F. are supported by Australian Government Research Training Program Scholarships. A.S.M. and A.M.F. are also supported by top-up scholarships funded by the Government of Western Australia. D.R.G. is supported by a DECRA fellowship (DE240100587). This material is based upon work supported by the Air Force Office of Scientific Research under award number FA2386-23-1-4081.
\end{acknowledgments}

\section*{Data availability}Data underlying the results presented in this paper are not publicly available at this time but may be obtained from the authors upon request. 

\appendix
\section{Estimating $r_0$ from the coupling fluctuations}\label{app:sr}Equation~(\ref{eq:SR}) gives the coupling into each mode averaged over turbulence. On millisecond timescales, the coupling fluctuates as the atmosphere changes. Here we use the variance of these fluctuations to estimate $r_0$. We square the overlap in Eq.~(\ref{eq:Psm}) before averaging over the turbulent phase $\varphi(\mathbf{u})$, as in Ref.~\cite{SRfibrecoupling}. This requires the fourth-order moment of the field, which in the phase-only model is
\begin{equation}
\Gamma_4=\left\langle e^{i[\varphi(\mathbf{u}_1)-\varphi(\mathbf{u}_2)+\varphi(\mathbf{u}_3)-\varphi(\mathbf{u}_4)]}\right\rangle
=\frac{B_{12}B_{23}B_{34}B_{41}}{B_{13}B_{24}},
\end{equation}
with $B_{ij}=B(\mathbf{u}_i-\mathbf{u}_j)$. The correlation between the couplings into modes $\mu$ and $\nu$ is then
\begin{multline}
\langle\eta_\mu\eta_\nu\rangle=\frac{1}{\mathcal{N}_\mu\mathcal{N}_\nu}\int\!\!\cdots\!\int\mathrm{d}^2u_1\cdots\mathrm{d}^2u_4\,\Gamma_4\\
\times G_\mu(\mathbf{u}_1)G_\mu^*(\mathbf{u}_2)G_\nu(\mathbf{u}_3)G_\nu^*(\mathbf{u}_4),
\label{eq:rho2SR}
\end{multline}
with $G_\mu$ and $\mathcal{N}_\mu$ as in Eq.~(\ref{eq:SR}). Equation~(\ref{eq:rho2SR}) describes the fluctuations of the single-mode and multi-mode signals $P_{\mathrm{opt}}$ and $Q_{\mathrm{opt}}$ of Eq.~(\ref{eq:Popt}), using the simulated coupler in Table~\ref{table:modes}.

The detected rates also fluctuate with scintillation, which changes the total flux but not the coupling. Scintillation multiplies $P_{\mathrm{opt}}$ and $Q_{\mathrm{opt}}$ by the same factor, so the detected single-mode fraction $P_{\mathrm{sm}}$ of Eq.~(\ref{eq:Psmdet}) is independent of scintillation. We sample Kolmogorov phase screens with the statistics of Eq.~(\ref{eq:rho2SR}) and compute $P_{\mathrm{sm}}$ for each value of $r_0$.

Using the computed values of $P_{\mathrm{sm}}$ we then simulate the distribution of the detected photons. We use only the point-source observations and divide each one into consecutive groups of $N$ photons. Given $N$, the number of single-mode photons in a group is binomially distributed, $n\sim\mathrm{Bin}(N,P_{\mathrm{sm}})$. We compare the simulated and measured values of the variance of $n/N$ to estimate $r_0$. This process assumes $P_{\mathrm{sm}}$ is constant within a group of $N$ photons, so we use $N=40$, or about 2~ms, which is much less than typical atmospheric coherence times. Scaling each estimate to zenith using the elevation of the source at the time of observation gives an estimated turbulence of $r_0=4.4\pm0.8$~cm at zenith for $\lambda=780$~nm.

\section{Type I and type II error rates}\label{app:errors}For a measurement of $n$ single-mode and $m$ multi-mode photons the likelihood ratio test statistic is
\begin{equation}
\Lambda_t=2\left[n\ln\frac{P_{\mathrm{sm}}^{(1)}}{P_{\mathrm{sm}}^{(0)}}+m\ln\frac{1-P_{\mathrm{sm}}^{(1)}}{1-P_{\mathrm{sm}}^{(0)}}\right].
\label{eq:app_lambda}
\end{equation}
We reject $H_0$ when $\Lambda_t>\tau_N$ and do not reject it when $\Lambda_t<\tau_N$. As $\Lambda_t$ only takes discrete values, when $\Lambda_t=\tau_N$ we reject $H_0$ with probability $\gamma_N$, where $\tau_N$ is the smallest threshold for which the probability of $\Lambda_t>\tau_N$ under $H_0$ is at most $\alpha$, and $\gamma_N$ is chosen so that the type I error rate equals $\alpha$. The type I and type II error rates are then
\begin{align}
\alpha={}&\sum_{n:\,\Lambda_t>\tau_N}P(n\mid H_0,N)\nonumber\\
&+\gamma_N\sum_{n:\,\Lambda_t=\tau_N}P(n\mid H_0,N),\\
\beta={}&\sum_{n:\,\Lambda_t<\tau_N}P(n\mid H_1,N)\nonumber\\
&+(1-\gamma_N)\sum_{n:\,\Lambda_t=\tau_N}P(n\mid H_1,N).
\end{align}

\bibliography{references}

\begin{thebibliography}{26}%
\makeatletter
\providecommand \@ifxundefined [1]{%
 \@ifx{#1\undefined}
}%
\providecommand \@ifnum [1]{%
 \ifnum #1\expandafter \@firstoftwo
 \else \expandafter \@secondoftwo
 \fi
}%
\providecommand \@ifx [1]{%
 \ifx #1\expandafter \@firstoftwo
 \else \expandafter \@secondoftwo
 \fi
}%
\providecommand \natexlab [1]{#1}%
\providecommand \enquote  [1]{``#1''}%
\providecommand \bibnamefont  [1]{#1}%
\providecommand \bibfnamefont [1]{#1}%
\providecommand \citenamefont [1]{#1}%
\providecommand \href@noop [0]{\@secondoftwo}%
\providecommand \href [0]{\begingroup \@sanitize@url \@href}%
\providecommand \@href[1]{\@@startlink{#1}\@@href}%
\providecommand \@@href[1]{\endgroup#1\@@endlink}%
\providecommand \@sanitize@url [0]{\catcode `\\12\catcode `\$12\catcode
  `\&12\catcode `\#12\catcode `\^12\catcode `\_12\catcode `\%12\relax}%
\providecommand \@@startlink[1]{}%
\providecommand \@@endlink[0]{}%
\providecommand \url  [0]{\begingroup\@sanitize@url \@url }%
\providecommand \@url [1]{\endgroup\@href {#1}{\urlprefix }}%
\providecommand \urlprefix  [0]{URL }%
\providecommand \Eprint [0]{\href }%
\providecommand \doibase [0]{https://doi.org/}%
\providecommand \selectlanguage [0]{\@gobble}%
\providecommand \bibinfo  [0]{\@secondoftwo}%
\providecommand \bibfield  [0]{\@secondoftwo}%
\providecommand \translation [1]{[#1]}%
\providecommand \BibitemOpen [0]{}%
\providecommand \bibitemStop [0]{}%
\providecommand \bibitemNoStop [0]{.\EOS\space}%
\providecommand \EOS [0]{\spacefactor3000\relax}%
\providecommand \BibitemShut  [1]{\csname bibitem#1\endcsname}%
\let\auto@bib@innerbib\@empty
\bibitem [{\citenamefont {Tsang}\ \emph {et~al.}(2016)\citenamefont {Tsang},
  \citenamefont {Nair},\ and\ \citenamefont {Lu}}]{TsangSuperresolution}%
  \BibitemOpen
  \bibfield  {author} {\bibinfo {author} {\bibfnamefont {M.}~\bibnamefont
  {Tsang}}, \bibinfo {author} {\bibfnamefont {R.}~\bibnamefont {Nair}},\ and\
  \bibinfo {author} {\bibfnamefont {X.-M.}\ \bibnamefont {Lu}},\ }\bibfield
  {title} {\bibinfo {title} {Quantum theory of superresolution for two
  incoherent optical point sources},\ }\href
  {https://doi.org/10.1103/PhysRevX.6.031033} {\bibfield  {journal} {\bibinfo
  {journal} {Phys. Rev. X}\ }\textbf {\bibinfo {volume} {6}},\ \bibinfo {pages}
  {031033} (\bibinfo {year} {2016})}\BibitemShut {NoStop}%
\bibitem [{\citenamefont {Lu}\ \emph {et~al.}(2018)\citenamefont {Lu},
  \citenamefont {Krovi}, \citenamefont {Nair}, \citenamefont {Guha},\ and\
  \citenamefont {Shapiro}}]{LuHypothesisTesting}%
  \BibitemOpen
  \bibfield  {author} {\bibinfo {author} {\bibfnamefont {X.-M.}\ \bibnamefont
  {Lu}}, \bibinfo {author} {\bibfnamefont {H.}~\bibnamefont {Krovi}}, \bibinfo
  {author} {\bibfnamefont {R.}~\bibnamefont {Nair}}, \bibinfo {author}
  {\bibfnamefont {S.}~\bibnamefont {Guha}},\ and\ \bibinfo {author}
  {\bibfnamefont {J.~H.}\ \bibnamefont {Shapiro}},\ }\bibfield  {title}
  {\bibinfo {title} {Quantum-optimal detection of one-versus-two incoherent
  optical sources with arbitrary separation},\ }\href@noop {} {\bibfield
  {journal} {\bibinfo  {journal} {npj Quantum Information}\ }\textbf {\bibinfo
  {volume} {4}},\ \bibinfo {pages} {64} (\bibinfo {year} {2018})}\BibitemShut
  {NoStop}%
\bibitem [{\citenamefont {Boucher}\ \emph {et~al.}(2020)\citenamefont
  {Boucher}, \citenamefont {Fabre}, \citenamefont {Labroille},\ and\
  \citenamefont {Treps}}]{BoucherMPLC}%
  \BibitemOpen
  \bibfield  {author} {\bibinfo {author} {\bibfnamefont {P.}~\bibnamefont
  {Boucher}}, \bibinfo {author} {\bibfnamefont {C.}~\bibnamefont {Fabre}},
  \bibinfo {author} {\bibfnamefont {G.}~\bibnamefont {Labroille}},\ and\
  \bibinfo {author} {\bibfnamefont {N.}~\bibnamefont {Treps}},\ }\bibfield
  {title} {\bibinfo {title} {Spatial optical mode demultiplexing as a practical
  tool for optimal transverse distance estimation},\ }\href
  {https://doi.org/10.1364/OPTICA.404746} {\bibfield  {journal} {\bibinfo
  {journal} {Optica}\ }\textbf {\bibinfo {volume} {7}},\ \bibinfo {pages}
  {1621} (\bibinfo {year} {2020})}\BibitemShut {NoStop}%
\bibitem [{\citenamefont {Santamaria}\ \emph {et~al.}(2024)\citenamefont
  {Santamaria}, \citenamefont {Sgobba},\ and\ \citenamefont
  {Lupo}}]{SantamariaMPLC}%
  \BibitemOpen
  \bibfield  {author} {\bibinfo {author} {\bibfnamefont {L.}~\bibnamefont
  {Santamaria}}, \bibinfo {author} {\bibfnamefont {F.}~\bibnamefont {Sgobba}},\
  and\ \bibinfo {author} {\bibfnamefont {C.}~\bibnamefont {Lupo}},\ }\bibfield
  {title} {\bibinfo {title} {Single-photon sub-rayleigh precision measurements
  of a pair of incoherent sources of unequal intensity},\ }\href
  {https://doi.org/10.1364/OPTICAQ.505457} {\bibfield  {journal} {\bibinfo
  {journal} {Optica Quantum}\ }\textbf {\bibinfo {volume} {2}},\ \bibinfo
  {pages} {46} (\bibinfo {year} {2024})}\BibitemShut {NoStop}%
\bibitem [{\citenamefont {Rouvi\`{e}re}\ \emph {et~al.}(2024)\citenamefont
  {Rouvi\`{e}re}, \citenamefont {Barral}, \citenamefont {Grateau},
  \citenamefont {Karuseichyk}, \citenamefont {Sorelli}, \citenamefont
  {Walschaers},\ and\ \citenamefont {Treps}}]{RouviereMPLC}%
  \BibitemOpen
  \bibfield  {author} {\bibinfo {author} {\bibfnamefont {C.}~\bibnamefont
  {Rouvi\`{e}re}}, \bibinfo {author} {\bibfnamefont {D.}~\bibnamefont
  {Barral}}, \bibinfo {author} {\bibfnamefont {A.}~\bibnamefont {Grateau}},
  \bibinfo {author} {\bibfnamefont {I.}~\bibnamefont {Karuseichyk}}, \bibinfo
  {author} {\bibfnamefont {G.}~\bibnamefont {Sorelli}}, \bibinfo {author}
  {\bibfnamefont {M.}~\bibnamefont {Walschaers}},\ and\ \bibinfo {author}
  {\bibfnamefont {N.}~\bibnamefont {Treps}},\ }\bibfield  {title} {\bibinfo
  {title} {Ultra-sensitive separation estimation of optical sources},\ }\href
  {https://doi.org/10.1364/OPTICA.500039} {\bibfield  {journal} {\bibinfo
  {journal} {Optica}\ }\textbf {\bibinfo {volume} {11}},\ \bibinfo {pages}
  {166} (\bibinfo {year} {2024})}\BibitemShut {NoStop}%
\bibitem [{\citenamefont {Tan}\ \emph {et~al.}(2023)\citenamefont {Tan},
  \citenamefont {Qi}, \citenamefont {Chen}, \citenamefont {Danner},
  \citenamefont {Kanchanawong},\ and\ \citenamefont {Tsang}}]{TanMPLC}%
  \BibitemOpen
  \bibfield  {author} {\bibinfo {author} {\bibfnamefont {X.-J.}\ \bibnamefont
  {Tan}}, \bibinfo {author} {\bibfnamefont {L.}~\bibnamefont {Qi}}, \bibinfo
  {author} {\bibfnamefont {L.}~\bibnamefont {Chen}}, \bibinfo {author}
  {\bibfnamefont {A.~J.}\ \bibnamefont {Danner}}, \bibinfo {author}
  {\bibfnamefont {P.}~\bibnamefont {Kanchanawong}},\ and\ \bibinfo {author}
  {\bibfnamefont {M.}~\bibnamefont {Tsang}},\ }\bibfield  {title} {\bibinfo
  {title} {Quantum-inspired superresolution for incoherent imaging},\ }\href
  {https://doi.org/10.1364/OPTICA.493227} {\bibfield  {journal} {\bibinfo
  {journal} {Optica}\ }\textbf {\bibinfo {volume} {10}},\ \bibinfo {pages}
  {1189} (\bibinfo {year} {2023})}\BibitemShut {NoStop}%
\bibitem [{\citenamefont {Wallis}\ \emph
  {et~al.}(2025{\natexlab{a}})\citenamefont {Wallis}, \citenamefont {Gozzard},
  \citenamefont {Frost}, \citenamefont {Collier}, \citenamefont {Maron},\ and\
  \citenamefont {Dix-Matthews}}]{WallisMPLC}%
  \BibitemOpen
  \bibfield  {author} {\bibinfo {author} {\bibfnamefont {J.~S.}\ \bibnamefont
  {Wallis}}, \bibinfo {author} {\bibfnamefont {D.~R.}\ \bibnamefont {Gozzard}},
  \bibinfo {author} {\bibfnamefont {A.~M.}\ \bibnamefont {Frost}}, \bibinfo
  {author} {\bibfnamefont {J.~J.}\ \bibnamefont {Collier}}, \bibinfo {author}
  {\bibfnamefont {N.}~\bibnamefont {Maron}},\ and\ \bibinfo {author}
  {\bibfnamefont {B.~P.}\ \bibnamefont {Dix-Matthews}},\ }\bibfield  {title}
  {\bibinfo {title} {Spatial mode demultiplexing for super-resolved source
  parameter estimation},\ }\href {https://doi.org/10.1364/OE.563503} {\bibfield
   {journal} {\bibinfo  {journal} {Opt. Express}\ }\textbf {\bibinfo {volume}
  {33}},\ \bibinfo {pages} {34651} (\bibinfo {year}
  {2025}{\natexlab{a}})}\BibitemShut {NoStop}%
\bibitem [{\citenamefont {Gozzard}\ \emph {et~al.}(2025)\citenamefont
  {Gozzard}, \citenamefont {Wallis}, \citenamefont {Frost}, \citenamefont
  {Collier}, \citenamefont {Maron}, \citenamefont {Dix-Matthews},\ and\
  \citenamefont {Vinsen}}]{MokuNN}%
  \BibitemOpen
  \bibfield  {author} {\bibinfo {author} {\bibfnamefont {D.~R.}\ \bibnamefont
  {Gozzard}}, \bibinfo {author} {\bibfnamefont {J.~S.}\ \bibnamefont {Wallis}},
  \bibinfo {author} {\bibfnamefont {A.~M.}\ \bibnamefont {Frost}}, \bibinfo
  {author} {\bibfnamefont {J.~J.}\ \bibnamefont {Collier}}, \bibinfo {author}
  {\bibfnamefont {N.}~\bibnamefont {Maron}}, \bibinfo {author} {\bibfnamefont
  {B.~P.}\ \bibnamefont {Dix-Matthews}},\ and\ \bibinfo {author} {\bibfnamefont
  {K.}~\bibnamefont {Vinsen}},\ }\bibfield  {title} {\bibinfo {title}
  {Super-resolution parameter estimation using machine learning-assisted
  spatial mode demultiplexing},\ }\bibfield  {journal} {\bibinfo  {journal}
  {Sensors}\ }\textbf {\bibinfo {volume} {25}},\ \href
  {https://doi.org/10.3390/s25175395} {10.3390/s25175395} (\bibinfo {year}
  {2025})\BibitemShut {NoStop}%
\bibitem [{\citenamefont {Santamaria}\ \emph {et~al.}(2023)\citenamefont
  {Santamaria}, \citenamefont {Pallotti}, \citenamefont {de~Cumis},
  \citenamefont {Dequal},\ and\ \citenamefont {Lupo}}]{LSantamaria:2023}%
  \BibitemOpen
  \bibfield  {author} {\bibinfo {author} {\bibfnamefont {L.}~\bibnamefont
  {Santamaria}}, \bibinfo {author} {\bibfnamefont {D.}~\bibnamefont
  {Pallotti}}, \bibinfo {author} {\bibfnamefont {M.~S.}\ \bibnamefont
  {de~Cumis}}, \bibinfo {author} {\bibfnamefont {D.}~\bibnamefont {Dequal}},\
  and\ \bibinfo {author} {\bibfnamefont {C.}~\bibnamefont {Lupo}},\ }\bibfield
  {title} {\bibinfo {title} {Spatial-mode demultiplexing for enhanced intensity
  and distance measurement},\ }\href {https://doi.org/10.1364/OE.486617}
  {\bibfield  {journal} {\bibinfo  {journal} {Opt. Express}\ }\textbf {\bibinfo
  {volume} {31}},\ \bibinfo {pages} {33930} (\bibinfo {year}
  {2023})}\BibitemShut {NoStop}%
\bibitem [{\citenamefont {Grateau}\ \emph {et~al.}(2026)\citenamefont
  {Grateau}, \citenamefont {Boeschoten}, \citenamefont {Favin-Lévêque},
  \citenamefont {Herrera},\ and\ \citenamefont {Treps}}]{AntoninTwoMPLCS}%
  \BibitemOpen
  \bibfield  {author} {\bibinfo {author} {\bibfnamefont {A.}~\bibnamefont
  {Grateau}}, \bibinfo {author} {\bibfnamefont {A.}~\bibnamefont {Boeschoten}},
  \bibinfo {author} {\bibfnamefont {T.}~\bibnamefont {Favin-Lévêque}},
  \bibinfo {author} {\bibfnamefont {I.}~\bibnamefont {Herrera}},\ and\ \bibinfo
  {author} {\bibfnamefont {N.}~\bibnamefont {Treps}},\ }\href
  {https://arxiv.org/abs/2601.14876} {\bibinfo {title} {Multiparameter
  estimation for the superresolution of two incoherent sources}} (\bibinfo
  {year} {2026}),\ \Eprint {https://arxiv.org/abs/2601.14876} {arXiv:2601.14876
  [quant-ph]} \BibitemShut {NoStop}%
\bibitem [{\citenamefont {Shringarpure}\ \emph {et~al.}(2026)\citenamefont
  {Shringarpure}, \citenamefont {Teo}, \citenamefont {Jeong}, \citenamefont
  {Evans}, \citenamefont {Sanchez-Soto}, \citenamefont {Grateau}, \citenamefont
  {Boeschoten},\ and\ \citenamefont {Treps}}]{BaesianMPLC}%
  \BibitemOpen
  \bibfield  {author} {\bibinfo {author} {\bibfnamefont {S.~U.}\ \bibnamefont
  {Shringarpure}}, \bibinfo {author} {\bibfnamefont {Y.~S.}\ \bibnamefont
  {Teo}}, \bibinfo {author} {\bibfnamefont {H.}~\bibnamefont {Jeong}}, \bibinfo
  {author} {\bibfnamefont {M.}~\bibnamefont {Evans}}, \bibinfo {author}
  {\bibfnamefont {L.~L.}\ \bibnamefont {Sanchez-Soto}}, \bibinfo {author}
  {\bibfnamefont {A.}~\bibnamefont {Grateau}}, \bibinfo {author} {\bibfnamefont
  {A.}~\bibnamefont {Boeschoten}},\ and\ \bibinfo {author} {\bibfnamefont
  {N.}~\bibnamefont {Treps}},\ }\href {https://arxiv.org/abs/2601.13972}
  {\bibinfo {title} {Experimental evidence-based sub-rayleigh source
  discrimination}} (\bibinfo {year} {2026}),\ \Eprint
  {https://arxiv.org/abs/2601.13972} {arXiv:2601.13972 [quant-ph]} \BibitemShut
  {NoStop}%
\bibitem [{\citenamefont {Kim}\ \emph {et~al.}(2025)\citenamefont {Kim},
  \citenamefont {Fitzgerald}, \citenamefont {Vievard}, \citenamefont {Lin},
  \citenamefont {Xin}, \citenamefont {Lucas}, \citenamefont {Guyon},
  \citenamefont {Lozi}, \citenamefont {Deo}, \citenamefont {Huby},
  \citenamefont {Lacour}, \citenamefont {Lallement}, \citenamefont
  {Amezcua-Correa}, \citenamefont {Leon-Saval}, \citenamefont {Norris},
  \citenamefont {Nowak}, \citenamefont {Sallum}, \citenamefont {Sarrazin},
  \citenamefont {Taras}, \citenamefont {Yerolatsitis},\ and\ \citenamefont
  {Jovanovic}}]{OnSkyPL}%
  \BibitemOpen
  \bibfield  {author} {\bibinfo {author} {\bibfnamefont {Y.~J.}\ \bibnamefont
  {Kim}}, \bibinfo {author} {\bibfnamefont {M.~P.}\ \bibnamefont {Fitzgerald}},
  \bibinfo {author} {\bibfnamefont {S.}~\bibnamefont {Vievard}}, \bibinfo
  {author} {\bibfnamefont {J.}~\bibnamefont {Lin}}, \bibinfo {author}
  {\bibfnamefont {Y.}~\bibnamefont {Xin}}, \bibinfo {author} {\bibfnamefont
  {M.}~\bibnamefont {Lucas}}, \bibinfo {author} {\bibfnamefont
  {O.}~\bibnamefont {Guyon}}, \bibinfo {author} {\bibfnamefont
  {J.}~\bibnamefont {Lozi}}, \bibinfo {author} {\bibfnamefont {V.}~\bibnamefont
  {Deo}}, \bibinfo {author} {\bibfnamefont {E.}~\bibnamefont {Huby}}, \bibinfo
  {author} {\bibfnamefont {S.}~\bibnamefont {Lacour}}, \bibinfo {author}
  {\bibfnamefont {M.}~\bibnamefont {Lallement}}, \bibinfo {author}
  {\bibfnamefont {R.}~\bibnamefont {Amezcua-Correa}}, \bibinfo {author}
  {\bibfnamefont {S.}~\bibnamefont {Leon-Saval}}, \bibinfo {author}
  {\bibfnamefont {B.}~\bibnamefont {Norris}}, \bibinfo {author} {\bibfnamefont
  {M.}~\bibnamefont {Nowak}}, \bibinfo {author} {\bibfnamefont
  {S.}~\bibnamefont {Sallum}}, \bibinfo {author} {\bibfnamefont
  {J.}~\bibnamefont {Sarrazin}}, \bibinfo {author} {\bibfnamefont
  {A.}~\bibnamefont {Taras}}, \bibinfo {author} {\bibfnamefont
  {S.}~\bibnamefont {Yerolatsitis}},\ and\ \bibinfo {author} {\bibfnamefont
  {N.}~\bibnamefont {Jovanovic}},\ }\bibfield  {title} {\bibinfo {title}
  {On-sky demonstration of subdiffraction-limited astronomical measurement
  using a photonic lantern},\ }\href {https://doi.org/10.3847/2041-8213/ae0739}
  {\bibfield  {journal} {\bibinfo  {journal} {The Astrophysical Journal
  Letters}\ }\textbf {\bibinfo {volume} {993}},\ \bibinfo {pages} {L3}
  (\bibinfo {year} {2025})}\BibitemShut {NoStop}%
\bibitem [{\citenamefont {Norris}\ \emph {et~al.}(2024)\citenamefont {Norris},
  \citenamefont {Leon-Saval}, \citenamefont {Wei}, \citenamefont {Betters},
  \citenamefont {Taras}, \citenamefont {Lin}, \citenamefont {Xin},
  \citenamefont {Kim}, \citenamefont {Fitzgerald}, \citenamefont {Sallum},
  \citenamefont {Sengupta}, \citenamefont {Gatkine}, \citenamefont {Jovanovic},
  \citenamefont {Mawet}, \citenamefont {Lozi}, \citenamefont {Vievard},
  \citenamefont {Deo}, \citenamefont {Lallement}, \citenamefont {Levinstein},\
  and\ \citenamefont {Guyon}}]{NorrisPL}%
  \BibitemOpen
  \bibfield  {author} {\bibinfo {author} {\bibfnamefont {B.~R.~M.}\
  \bibnamefont {Norris}}, \bibinfo {author} {\bibfnamefont {S.~G.}\
  \bibnamefont {Leon-Saval}}, \bibinfo {author} {\bibfnamefont
  {J.}~\bibnamefont {Wei}}, \bibinfo {author} {\bibfnamefont {C.~H.}\
  \bibnamefont {Betters}}, \bibinfo {author} {\bibfnamefont {A.}~\bibnamefont
  {Taras}}, \bibinfo {author} {\bibfnamefont {J.}~\bibnamefont {Lin}}, \bibinfo
  {author} {\bibfnamefont {Y.}~\bibnamefont {Xin}}, \bibinfo {author}
  {\bibfnamefont {Y.~J.}\ \bibnamefont {Kim}}, \bibinfo {author} {\bibfnamefont
  {M.}~\bibnamefont {Fitzgerald}}, \bibinfo {author} {\bibfnamefont
  {S.}~\bibnamefont {Sallum}}, \bibinfo {author} {\bibfnamefont
  {A.}~\bibnamefont {Sengupta}}, \bibinfo {author} {\bibfnamefont
  {P.}~\bibnamefont {Gatkine}}, \bibinfo {author} {\bibfnamefont
  {N.}~\bibnamefont {Jovanovic}}, \bibinfo {author} {\bibfnamefont
  {D.}~\bibnamefont {Mawet}}, \bibinfo {author} {\bibfnamefont
  {J.}~\bibnamefont {Lozi}}, \bibinfo {author} {\bibfnamefont {S.}~\bibnamefont
  {Vievard}}, \bibinfo {author} {\bibfnamefont {V.}~\bibnamefont {Deo}},
  \bibinfo {author} {\bibfnamefont {M.}~\bibnamefont {Lallement}}, \bibinfo
  {author} {\bibfnamefont {D.}~\bibnamefont {Levinstein}},\ and\ \bibinfo
  {author} {\bibfnamefont {O.}~\bibnamefont {Guyon}},\ }\bibfield  {title}
  {\bibinfo {title} {{The photonic lantern wavefront sensor and imager: focal
  plane wavefront sensing and optimal imaging at the diffraction limit and
  beyond}},\ }in\ \href {https://doi.org/10.1117/12.3019643} {\emph {\bibinfo
  {booktitle} {Adaptive Optics Systems IX}}},\ Vol.\ \bibinfo {volume}
  {13097},\ \bibinfo {editor} {edited by\ \bibinfo {editor} {\bibfnamefont
  {K.~J.}\ \bibnamefont {Jackson}}, \bibinfo {editor} {\bibfnamefont
  {D.}~\bibnamefont {Schmidt}},\ and\ \bibinfo {editor} {\bibfnamefont
  {E.}~\bibnamefont {Vernet}}},\ \bibinfo {organization} {International Society
  for Optics and Photonics}\ (\bibinfo  {publisher} {SPIE},\ \bibinfo {year}
  {2024})\ p.\ \bibinfo {pages} {130971I}\BibitemShut {NoStop}%
\bibitem [{\citenamefont {Pushkina}\ \emph {et~al.}(2021)\citenamefont
  {Pushkina}, \citenamefont {Maltese}, \citenamefont {Costa-Filho},
  \citenamefont {Patel},\ and\ \citenamefont {Lvovsky}}]{PuskhinaSLM}%
  \BibitemOpen
  \bibfield  {author} {\bibinfo {author} {\bibfnamefont {A.}~\bibnamefont
  {Pushkina}}, \bibinfo {author} {\bibfnamefont {G.}~\bibnamefont {Maltese}},
  \bibinfo {author} {\bibfnamefont {J.}~\bibnamefont {Costa-Filho}}, \bibinfo
  {author} {\bibfnamefont {P.}~\bibnamefont {Patel}},\ and\ \bibinfo {author}
  {\bibfnamefont {A.}~\bibnamefont {Lvovsky}},\ }\bibfield  {title} {\bibinfo
  {title} {Superresolution linear optical imaging in the far field},\
  }\href@noop {} {\bibfield  {journal} {\bibinfo  {journal} {Phys. Rev. Lett.}\
  }\textbf {\bibinfo {volume} {127}},\ \bibinfo {pages} {253602} (\bibinfo
  {year} {2021})}\BibitemShut {NoStop}%
\bibitem [{\citenamefont {Yang}\ \emph {et~al.}(2016)\citenamefont {Yang},
  \citenamefont {Tashchilina}, \citenamefont {Moiseev}, \citenamefont {Simon},\
  and\ \citenamefont {Lvovsky}}]{FYangLO}%
  \BibitemOpen
  \bibfield  {author} {\bibinfo {author} {\bibfnamefont {F.}~\bibnamefont
  {Yang}}, \bibinfo {author} {\bibfnamefont {A.}~\bibnamefont {Tashchilina}},
  \bibinfo {author} {\bibfnamefont {E.~S.}\ \bibnamefont {Moiseev}}, \bibinfo
  {author} {\bibfnamefont {C.}~\bibnamefont {Simon}},\ and\ \bibinfo {author}
  {\bibfnamefont {A.~I.}\ \bibnamefont {Lvovsky}},\ }\bibfield  {title}
  {\bibinfo {title} {Far-field linear optical superresolution via heterodyne
  detection in a higher-order local oscillator mode},\ }\href
  {https://doi.org/10.1364/OPTICA.3.001148} {\bibfield  {journal} {\bibinfo
  {journal} {Optica}\ }\textbf {\bibinfo {volume} {3}},\ \bibinfo {pages}
  {1148} (\bibinfo {year} {2016})}\BibitemShut {NoStop}%
\bibitem [{\citenamefont {Huang}\ and\ \citenamefont
  {Lupo}(2021)}]{NoiselessHT}%
  \BibitemOpen
  \bibfield  {author} {\bibinfo {author} {\bibfnamefont {Z.}~\bibnamefont
  {Huang}}\ and\ \bibinfo {author} {\bibfnamefont {C.}~\bibnamefont {Lupo}},\
  }\bibfield  {title} {\bibinfo {title} {Quantum hypothesis testing for
  exoplanet detection},\ }\href
  {https://doi.org/10.1103/PhysRevLett.127.130502} {\bibfield  {journal}
  {\bibinfo  {journal} {Phys. Rev. Lett.}\ }\textbf {\bibinfo {volume} {127}},\
  \bibinfo {pages} {130502} (\bibinfo {year} {2021})}\BibitemShut {NoStop}%
\bibitem [{\citenamefont {Schlichtholz}\ \emph {et~al.}(2024)\citenamefont
  {Schlichtholz}, \citenamefont {Linowski}, \citenamefont {Walschaers},
  \citenamefont {Treps}, \citenamefont {Rudnicki},\ and\ \citenamefont
  {Sorelli}}]{CrossTalkHT}%
  \BibitemOpen
  \bibfield  {author} {\bibinfo {author} {\bibfnamefont {K.}~\bibnamefont
  {Schlichtholz}}, \bibinfo {author} {\bibfnamefont {T.}~\bibnamefont
  {Linowski}}, \bibinfo {author} {\bibfnamefont {M.}~\bibnamefont
  {Walschaers}}, \bibinfo {author} {\bibfnamefont {N.}~\bibnamefont {Treps}},
  \bibinfo {author} {\bibfnamefont {L.}~\bibnamefont {Rudnicki}},\ and\
  \bibinfo {author} {\bibfnamefont {G.}~\bibnamefont {Sorelli}},\ }\bibfield
  {title} {\bibinfo {title} {Practical tests for sub-rayleigh source
  discriminations with imperfect demultiplexers},\ }\href
  {https://doi.org/10.1364/OPTICAQ.502459} {\bibfield  {journal} {\bibinfo
  {journal} {Optica Quantum}\ }\textbf {\bibinfo {volume} {2}},\ \bibinfo
  {pages} {29} (\bibinfo {year} {2024})}\BibitemShut {NoStop}%
\bibitem [{\citenamefont {Linowski}\ \emph {et~al.}(2025)\citenamefont
  {Linowski}, \citenamefont {Schlichtholz},\ and\ \citenamefont
  {Sorelli}}]{PracticalHT}%
  \BibitemOpen
  \bibfield  {author} {\bibinfo {author} {\bibfnamefont {T.}~\bibnamefont
  {Linowski}}, \bibinfo {author} {\bibfnamefont {K.}~\bibnamefont
  {Schlichtholz}},\ and\ \bibinfo {author} {\bibfnamefont {G.}~\bibnamefont
  {Sorelli}},\ }\bibfield  {title} {\bibinfo {title} {Quantum-inspired
  exoplanet detection in the presence of experimental imperfections},\ }\href
  {https://doi.org/10.1103/qq99-jmpv} {\bibfield  {journal} {\bibinfo
  {journal} {Phys. Rev. Appl.}\ }\textbf {\bibinfo {volume} {24}},\ \bibinfo
  {pages} {044052} (\bibinfo {year} {2025})}\BibitemShut {NoStop}%
\bibitem [{\citenamefont {Wallis}\ \emph
  {et~al.}(2025{\natexlab{b}})\citenamefont {Wallis}, \citenamefont {Gozzard},
  \citenamefont {Frost}, \citenamefont {Dix-Matthews}, \citenamefont {Maron},\
  and\ \citenamefont {Collier}}]{WallisDCF}%
  \BibitemOpen
  \bibfield  {author} {\bibinfo {author} {\bibfnamefont {J.~S.}\ \bibnamefont
  {Wallis}}, \bibinfo {author} {\bibfnamefont {D.~R.}\ \bibnamefont {Gozzard}},
  \bibinfo {author} {\bibfnamefont {A.~M.}\ \bibnamefont {Frost}}, \bibinfo
  {author} {\bibfnamefont {B.~P.}\ \bibnamefont {Dix-Matthews}}, \bibinfo
  {author} {\bibfnamefont {N.}~\bibnamefont {Maron}},\ and\ \bibinfo {author}
  {\bibfnamefont {J.~J.}\ \bibnamefont {Collier}},\ }\bibfield  {title}
  {\bibinfo {title} {Super-resolving binary-source hypothesis testing with a
  double-clad fiber coupler},\ }\href {https://doi.org/10.1364/OL.573054}
  {\bibfield  {journal} {\bibinfo  {journal} {Opt. Lett.}\ }\textbf {\bibinfo
  {volume} {50}},\ \bibinfo {pages} {7191} (\bibinfo {year}
  {2025}{\natexlab{b}})}\BibitemShut {NoStop}%
\bibitem [{\citenamefont {Shaklan}\ and\ \citenamefont
  {Roddier}(1988)}]{SRfibrecoupling}%
  \BibitemOpen
  \bibfield  {author} {\bibinfo {author} {\bibfnamefont {S.}~\bibnamefont
  {Shaklan}}\ and\ \bibinfo {author} {\bibfnamefont {F.}~\bibnamefont
  {Roddier}},\ }\bibfield  {title} {\bibinfo {title} {Coupling starlight into
  single-mode fiber optics},\ }\href {https://doi.org/10.1364/AO.27.002334}
  {\bibfield  {journal} {\bibinfo  {journal} {Appl. Opt.}\ }\textbf {\bibinfo
  {volume} {27}},\ \bibinfo {pages} {2334} (\bibinfo {year}
  {1988})}\BibitemShut {NoStop}%
\bibitem [{\citenamefont {Burdge}\ \emph {et~al.}(2024)\citenamefont {Burdge},
  \citenamefont {El-Badry}, \citenamefont {Kara}, \citenamefont {Canizares},
  \citenamefont {Chakrabarty}, \citenamefont {Frebel}, \citenamefont
  {Millholland}, \citenamefont {Rappaport}, \citenamefont {Simcoe},\ and\
  \citenamefont {Vanderburg}}]{BlackholeDiscovery}%
  \BibitemOpen
  \bibfield  {author} {\bibinfo {author} {\bibfnamefont {K.~B.}\ \bibnamefont
  {Burdge}}, \bibinfo {author} {\bibfnamefont {K.}~\bibnamefont {El-Badry}},
  \bibinfo {author} {\bibfnamefont {E.}~\bibnamefont {Kara}}, \bibinfo {author}
  {\bibfnamefont {C.}~\bibnamefont {Canizares}}, \bibinfo {author}
  {\bibfnamefont {D.}~\bibnamefont {Chakrabarty}}, \bibinfo {author}
  {\bibfnamefont {A.}~\bibnamefont {Frebel}}, \bibinfo {author} {\bibfnamefont
  {S.~C.}\ \bibnamefont {Millholland}}, \bibinfo {author} {\bibfnamefont
  {S.}~\bibnamefont {Rappaport}}, \bibinfo {author} {\bibfnamefont
  {R.}~\bibnamefont {Simcoe}},\ and\ \bibinfo {author} {\bibfnamefont
  {A.}~\bibnamefont {Vanderburg}},\ }\bibfield  {title} {\bibinfo {title} {The
  black hole low-mass x-ray binary v404 cygni is part of a wide triple},\
  }\href {https://doi.org/10.1038/s41586-024-08120-6} {\bibfield  {journal}
  {\bibinfo  {journal} {Nature}\ }\textbf {\bibinfo {volume} {635}},\ \bibinfo
  {pages} {316} (\bibinfo {year} {2024})}\BibitemShut {NoStop}%
\bibitem [{\citenamefont {Lemire-Renaud}\ \emph {et~al.}(2010)\citenamefont
  {Lemire-Renaud}, \citenamefont {Rivard}, \citenamefont {Strupler},
  \citenamefont {Morneau}, \citenamefont {Verpillat}, \citenamefont {Daxhelet},
  \citenamefont {Godbout},\ and\ \citenamefont {Boudoux}}]{LemireRenaudDCFC}%
  \BibitemOpen
  \bibfield  {author} {\bibinfo {author} {\bibfnamefont {S.}~\bibnamefont
  {Lemire-Renaud}}, \bibinfo {author} {\bibfnamefont {M.}~\bibnamefont
  {Rivard}}, \bibinfo {author} {\bibfnamefont {M.}~\bibnamefont {Strupler}},
  \bibinfo {author} {\bibfnamefont {D.}~\bibnamefont {Morneau}}, \bibinfo
  {author} {\bibfnamefont {F.}~\bibnamefont {Verpillat}}, \bibinfo {author}
  {\bibfnamefont {X.}~\bibnamefont {Daxhelet}}, \bibinfo {author}
  {\bibfnamefont {N.}~\bibnamefont {Godbout}},\ and\ \bibinfo {author}
  {\bibfnamefont {C.}~\bibnamefont {Boudoux}},\ }\bibfield  {title} {\bibinfo
  {title} {Double-clad fiber coupler for endoscopy},\ }\href
  {https://doi.org/10.1364/OE.18.009755} {\bibfield  {journal} {\bibinfo
  {journal} {Opt. Express}\ }\textbf {\bibinfo {volume} {18}},\ \bibinfo
  {pages} {9755} (\bibinfo {year} {2010})}\BibitemShut {NoStop}%
\bibitem [{\citenamefont {Xin}\ \emph {et~al.}(2022)\citenamefont {Xin} \emph
  {et~al.}}]{XinPLN}%
  \BibitemOpen
  \bibfield  {author} {\bibinfo {author} {\bibfnamefont {Y.}~\bibnamefont
  {Xin}} \emph {et~al.},\ }\bibfield  {title} {\bibinfo {title} {Efficient
  detection and characterization of exoplanets within the diffraction limit:
  nulling with a mode-selective photonic lantern},\ }\href@noop {} {\bibfield
  {journal} {\bibinfo  {journal} {The Astrophysical Journal}\ }\textbf
  {\bibinfo {volume} {938}} (\bibinfo {year} {2022})},\ \Eprint
  {https://arxiv.org/abs/2209.07644} {arXiv:2209.07644} \BibitemShut {NoStop}%
\bibitem [{\citenamefont {Rigaut}\ \emph {et~al.}(1998)\citenamefont {Rigaut},
  \citenamefont {V{\'e}ran},\ and\ \citenamefont {Lai}}]{RigautAO}%
  \BibitemOpen
  \bibfield  {author} {\bibinfo {author} {\bibfnamefont {F.~J.}\ \bibnamefont
  {Rigaut}}, \bibinfo {author} {\bibfnamefont {J.-P.}\ \bibnamefont
  {V{\'e}ran}},\ and\ \bibinfo {author} {\bibfnamefont {O.}~\bibnamefont
  {Lai}},\ }\bibfield  {title} {\bibinfo {title} {Analytical model for
  shack-hartmann-based adaptive optics systems},\ }in\ \href
  {https://doi.org/10.1117/12.321649} {\emph {\bibinfo {booktitle} {Adaptive
  Optical System Technologies}}},\ \bibinfo {series} {Proc. SPIE}, Vol.\
  \bibinfo {volume} {3353}\ (\bibinfo {year} {1998})\ pp.\ \bibinfo {pages}
  {1038--1048}\BibitemShut {NoStop}%
\bibitem [{\citenamefont {Jolissaint}\ \emph {et~al.}(2006)\citenamefont
  {Jolissaint}, \citenamefont {V{\'e}ran},\ and\ \citenamefont
  {Conan}}]{JolissaintAO}%
  \BibitemOpen
  \bibfield  {author} {\bibinfo {author} {\bibfnamefont {L.}~\bibnamefont
  {Jolissaint}}, \bibinfo {author} {\bibfnamefont {J.-P.}\ \bibnamefont
  {V{\'e}ran}},\ and\ \bibinfo {author} {\bibfnamefont {R.}~\bibnamefont
  {Conan}},\ }\bibfield  {title} {\bibinfo {title} {Analytical modeling of
  adaptive optics: foundations of the phase spatial power spectrum approach},\
  }\href@noop {} {\bibfield  {journal} {\bibinfo  {journal} {Journal of the
  Optical Society of America A}\ }\textbf {\bibinfo {volume} {23}},\ \bibinfo
  {pages} {382} (\bibinfo {year} {2006})}\BibitemShut {NoStop}%
\bibitem [{\citenamefont {Becerra-Deana}\ \emph {et~al.}(2025)\citenamefont
  {Becerra-Deana}, \citenamefont {{Poinsinet de Sivry-Houle}}, \citenamefont
  {Virally}, \citenamefont {Boudoux},\ and\ \citenamefont
  {Godbout}}]{BecerraDeanaDCFLantern}%
  \BibitemOpen
  \bibfield  {author} {\bibinfo {author} {\bibfnamefont {R.~I.}\ \bibnamefont
  {Becerra-Deana}}, \bibinfo {author} {\bibfnamefont {M.}~\bibnamefont
  {{Poinsinet de Sivry-Houle}}}, \bibinfo {author} {\bibfnamefont
  {S.}~\bibnamefont {Virally}}, \bibinfo {author} {\bibfnamefont
  {C.}~\bibnamefont {Boudoux}},\ and\ \bibinfo {author} {\bibfnamefont
  {N.}~\bibnamefont {Godbout}},\ }\bibfield  {title} {\bibinfo {title}
  {Mode-selective photonic lanterns with double-clad fibers},\ }\href@noop {}
  {\bibfield  {journal} {\bibinfo  {journal} {Journal of Lightwave Technology}\
  }\textbf {\bibinfo {volume} {43}},\ \bibinfo {pages} {5829} (\bibinfo {year}
  {2025})},\ \Eprint {https://arxiv.org/abs/2410.23403} {arXiv:2410.23403}
  \BibitemShut {NoStop}%
\end{thebibliography}%

\end{document}